\documentclass[entropy,article,accept,moreauthors]{mdpi} 
\usepackage{bbold}
\usepackage{braket}
\usetikzlibrary{patterns,decorations.pathreplacing,calligraphy}
\usepackage{color}
\usepackage{xcolor}
\usepackage{MnSymbol}
\firstpage{1} 
\pubvolume{27}
\issuenum{1}
\articlenumber{407}
\pubyear{2025}
\copyrightyear{2025}
\datereceived{ 8 March 2025} 
\daterevised{ 6 April 2025} % Comment out if no revised date
\dateaccepted{ 7 April 2025} 
\datepublished{ 10 April 2025} 
\hreflink{https://doi.org/10.3390/e27040407} % If needed use \linebreak
\newcommand{\titleinfo}{Measurement-induced symmetry restoration and quantum Mpemba effect}
\Title{\titleinfo}

\TitleCitation{\titleinfo}

\Author{Giuseppe Di Giulio$^{1}$\orcidB, Xhek Turkeshi $^{2}$\orcidA{}, Sara Murciano $^{3}$\orcidC{}}

\AuthorNames{Giuseppe Di Giulio, Xhek Turkeshi, Sara Murciano }

\AuthorCitation{ Di Giulio, G.; Turkeshi, X.; Murciano, S.}
\address{%
${1}$ Oscar Klein Centre and Department of Physics, Stockholm University, AlbaNova, 106 91 Stockholm, Sweden
\\%
${2}$ Institut f\"ur Theoretische Physik, Universit\"at zu K\"oln, Z\"ulpicher Strasse 77a, 50937 K\"oln, Germany
\\
${3}$ Universit\'e Paris-Saclay, CNRS, LPTMS, 91405, Orsay, France
}

\abstract{
Monitoring a quantum system can profoundly alter its dynamical properties, leading to nontrivial emergent phenomena. 
In this work, we demonstrate that dynamical measurements strongly influence the evolution of symmetry in many-body quantum systems. 
Specifically, we demonstrate that monitored systems governed by non-Hermitian dynamics exhibit a quantum Mpemba effect, where systems with stronger initial asymmetry relax faster to a symmetric state. 
Crucially, this phenomenon is purely measurement-induced: in the absence of measurements, we find states where the corresponding unitary evolution does not display any Mpemba effect. 
Furthermore, we uncover a novel measurement-induced symmetry restoration mechanism: below a critical measurement rate, the symmetry remains broken, but beyond a threshold, it is fully restored in the thermodynamic limit—along with the emergence of the quantum Mpemba effect.
}


\begin{document}

%%%%%%%%%%%%%%%%%%%%%%%%%%%%%%%%%%%%%%%%%%
\section{Introduction}

The Mpemba effect is a counterintuitive phenomenon in statistical physics where a system initially at a higher temperature relaxes faster to equilibrium than a cooler one. 
First observed in classical fluids undergoing thermal quenches, has been later reported in a variety of classical systems~\cite{Mpemba1979}, from colloids \cite{pnas,kumar2020exponentially} to granular gases~\cite{lasanta2017when} (see Ref. \cite{Teza2025} for a detailed review about this phenomenon and possible explanations of the underlying mechanisms in classical systems). 
More recently, the Mpemba effect has been extended in the quantum domain. This so-called quantum Mpemba effect describes situations in which a system farther from equilibrium relaxes faster than one closer to it. 
In closed quantum systems, the effect was first identified through the study of symmetry-breaking states and their relaxation under unitary dynamics \cite{Ares:2022koq}. The underlying mechanism has been particularly well understood in integrable models, where quasiparticle dynamics govern relaxation \cite{Rylands:2023yzx,Chalas:2024wjz,Murciano:2023qrv,Rylands:2024fio,Yamashika:2024mut,Yamashika:2024hpr,Klobas:2024mlb}. In these systems, the effect occurs when the more asymmetric initial state contains faster quasiparticles, which accelerate the symmetry restoration.

While integrability offers a valuable framework, the Mpemba effect is not restricted to integrable dynamics, raising the question of whether a more general criterion for its occurrence can be established. Indeed, beyond integrable systems, the quantum Mpemba effect has been observed in a variety of non-integrable settings, including random circuits \cite{Turkeshi:2024juo,liu2024symmetry}, dual-unitary models \cite{Foligno:2024jpq}, many-body localized systems \cite{liu2024quantummpembaeffectsmanybody}, and trapped ion experiments \cite{Joshi:2024sup}. It has also been explored in open quantum systems, where dissipation, decoherence, and irreversible dynamics introduce additional complexity \cite{nava2019,Chatterjee2023,Chatterjee2023multiple,shapirainverse2024,Zatsarynna:2025efg,nava2024mpemba}. Various setups, such as systems with weak gain/loss dissipation \cite{Ares:2024nkh,Caceffo:2024jbc}, Markovian and non-Markovian baths \cite{carollo2021,Zhang:2024juc,moroderThermodynamics2024}, have provided further insights into the conditions under which the effect manifests (see the review \cite{Ares:2025onj} for a detailed list of references about the quantum version of the phenomenon in open systems). 

A distinct regime lies between closed and open quantum systems: monitored quantum systems, where unitary evolution competes with quantum measurements~\cite{fisher2023randomquantumcircuits,potter2022quantumsciencesandtechnology,fazio2024manybodyopenquantumsystems,Lunt2022}. Unlike open systems, measurements preserve the purity of the state, hence they do not globally thermalize as open systems do. 
Instead, measurements introduce a stochastic many-body process, fundamentally altering the system’s dynamical properties~\cite{jacobs2014quantummeasurementtheory}. 
Crucially, these phenomena are genuinely encoded in individual quantum trajectories that are determined, or post-selected, by the measurement outcomes~\cite{cao2019entanglementina,sierant2022dissipativefloquetdynamics,gullans2020scalableprobesof}. 
The cornerstone example is measurement-induced phase transitions, where quantum information properties undergo abrupt structural changes as the measurement rate increases~\cite{li2019measurementdrivenentanglement,skinner2019measurementinducedphase,bao2020theoryofthe,gullans2020dynamicalpurificationphase,jian2020measurementinducedcriticality,zabalo202criticalpropertiesof,zabalo2022operatorscalingdimensions,lirasolanilla2024multipartiteentanglementstructuremonitored,sierant2022universalbehaviorbeyond,sierant2022measurementinducedphase,nahum2020entanglement}.

In this work, we demonstrate that continuously monitoring a many-body Hamiltonian system~\cite{alberton2021entanglement,turkeshi2021measurementinducedentanglement,turkeshi2022entanglementtransitionsfrom,tirrito2023full,lumia,piccitto2022entanglement,Piccitto_2024,Tsitsishvili2024measurement,muzzi2025entanglement,coppola2022growth,cecile2024measurementinducedphasetransitionsmatrix,loio2023purification,graham2023topological,nehra2024controllingmeasurementinducedphase,buchhold2022revealingmeasurementinducedphasetransitions,starchl2025generalizedzenoeffectentanglement,
PhysRevB.109.L060302,difresco2025entanglementgrowthdarkintervals} can induce the quantum Mpemba effect even in setups when the purely unitary, unmonitored system does not present the Mpemba phenomenology. 
Concretely, we consider the post-selected trajectory with no quantum jumps -- the so-called no-click limit~\cite{lee2014heralded,biella2021manybodyquantumzeno}, obtaining an effective non-Hermitian evolution which serves as a powerful framework for studying quantum systems interacting with an environment~\cite{turkeshi2023entanglementandcorrelation,legal2023volume,zerba2023measurement,paviglianiti2023multipartite,guerra2025correlationskrylovspreadnonhermitian,soares2024nonunitary,zhang2023antiunitarysymmetrybreakinghierarchy,granet2023volumelaw,moca2020quantum,dora2022correlations,dora2022full,despres2024breakdown,dora2023quantum,bacsi2021dynamics,dupays2025slow,Chakrabarti:2025hsb}. 

\begin{figure}[t!]
\centering
\includegraphics[width=1.0\textwidth]{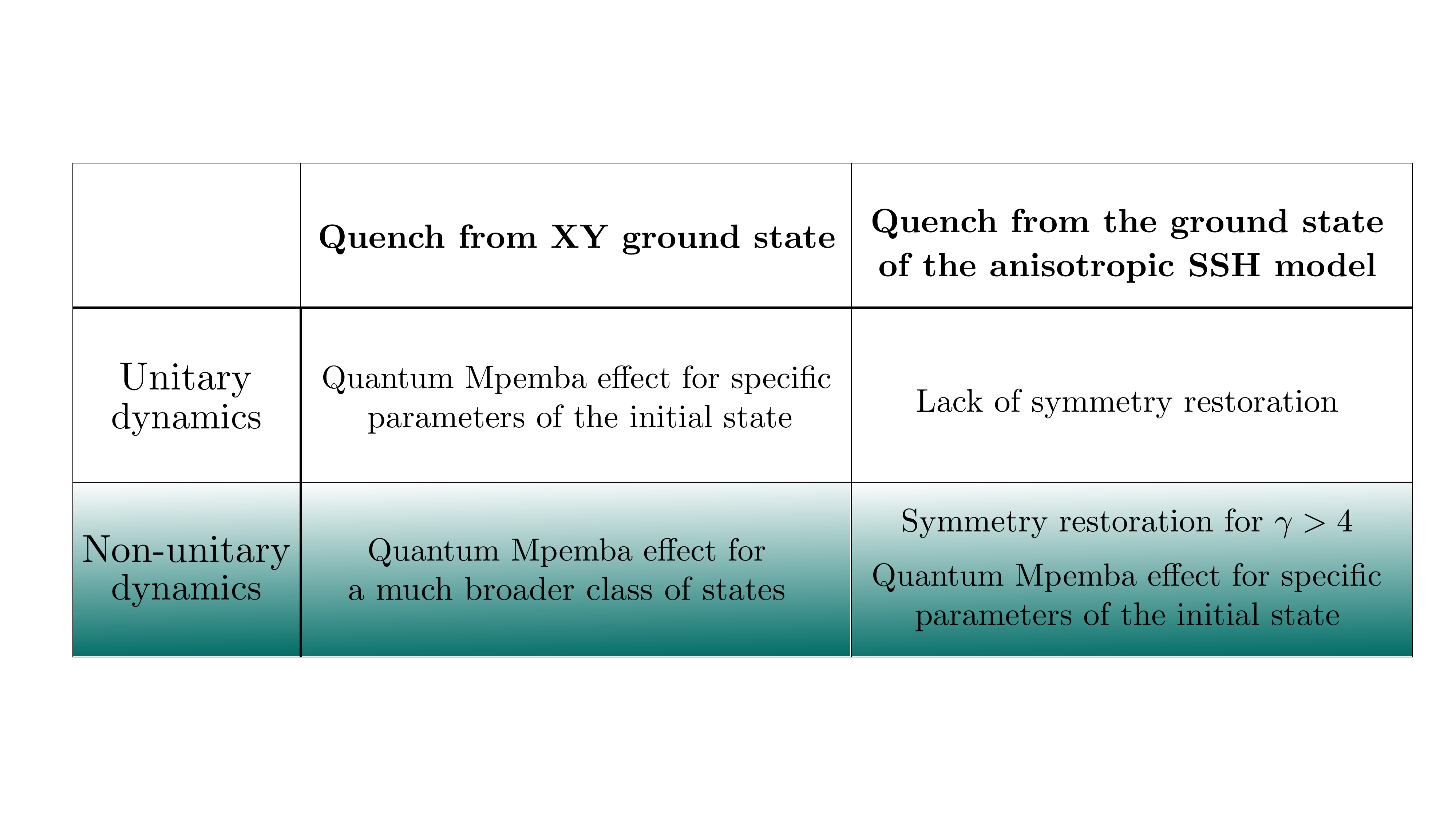}
\vspace{-1.5cm}
    \caption{
    Comparison of the presence of symmetry restoration and quantum Mpemba effect under purely unitary (first row) and non-unitary (second row) evolution for the two quantum quenches considered in this work.
   }
    \label{fig:table}
\end{figure}

We consider two distinct setups:
\begin{enumerate}
    \item \textit{Monitored hopping fermion dynamics from the $XY$ ground state}. The system is initially prepared in the ground state of the $XY$ spin chain and evolves under a $U(1)$-preserving Hamiltonian. While previous studies have established conditions for the quantum Mpemba effect in purely unitary evolution \cite{Murciano:2023qrv} or with weak dissipation \cite{Ares:2024nkh}, our results reveal that monitoring significantly broadens the class of initial states exhibiting this phenomenon (see left column of the table in Fig.\,\ref{fig:table}). This provides a clear demonstration of a genuinely measurement-induced Mpemba effect—one that arises exclusively due to the interplay between unitary dynamics and measurement. 
    \item \textit{Quench from an anisotropic Su–Schrieffer–Heeger (SSH)  ground state}. The system starts from the ground state of an anisotropic SSH model and evolves under the same $U(1)$-preserving Hamiltonian. Under unitary evolution, the symmetry remains broken, preventing the Mpemba effect from emerging. Strikingly, beyond a critical measurement rate, the symmetry is fully restored, triggering the Mpemba effect (see right column of the table in Fig.\,\ref{fig:table}). This constitutes a genuinely measurement-induced symmetry restoration, with no counterpart in purely unitary dynamics.
\end{enumerate} 
Our results demonstrate that measurement fundamentally reshapes relaxation dynamics, giving rise to new instances of the Mpemba effect absent in both unitary and dissipative settings. 
This work reveals the critical role quantum measurements play in symmetry restoration and non-equilibrium dynamics.
We would like to emphasize that the setting studied in this work is quite different from previous studies on the quantum Mpemba effect in open systems. In earlier works, this phenomenon was observed after a thermal quench \cite{nava2019} or in Markovian open quantum systems \cite{carollo2021}, where the evolution is governed by a Lindblad equation that describes the system’s interaction with the environment. In those cases, the quantum Mpemba effect refers to an unusual dynamical process in which a system initially further from equilibrium relaxes faster than one closer to equilibrium.
In contrast, our approach indirectly tracks how quickly equilibrium is reached by focusing on a symmetry that is initially broken and then locally restored over time. We define the quantum Mpemba effect in this context as the scenario where a stronger initial symmetry breaking leads to a faster restoration after a global quench. Additionally, the evolution of our post-measurement state follows a (stochastic) Schr\"odinger equation, further distinguishing our setup from previous studies.

\section{Methods}
\label{sec:entanglementAsy}
The primary tool we use to explore the Mpemba phenomenology is the entanglement asymmetry, which is closely tied to the resource theory of asymmetry~\cite{chitambar2019quantumresourcetheories}. 
Before introducing the models of interest, we briefly review its definition and key properties.
Throughout this work, we consider a dynamical framework with a $U(1)$ symmetry generated by the charge operator $Q=\sum_{j}c^{\dagger}_j c_j$, where $c_j$ and $c^\dagger_j$ are fermionic operators. A pure state $\ket{\psi}$ has a well-defined charge if and only if it is an eigenstate of $Q$. Otherwise, the state is said to be \emph{asymmetric} or \emph{symmetry-broken}.  
For any given bipartition $A\cup B$ and an asymmetric state $\ket{\psi}$, the reduced density matrix $\rho_A=\mathrm{Tr}_B(|\psi\rangle\langle\psi|)$ does not commute with the local charge operator $Q_A \equiv \sum_{j\in A}c^{\dagger}_j c_j$. 
To quantify the degree of symmetry breaking in subsystem $A$, we define the \emph{decohered}, or \textit{symmetrized}, density matrix  $\rho_{A,Q}=\sum_q \Pi_q\rho_A\Pi_q$, where $\Pi_q$ is the projector onto the eigenspace of $Q_A$ with charge $q$. By construction, $\rho_{A,Q}$ satisfies $[\rho_{A,Q},Q_A]=0$, regardless of the symmetry properties of the global state $\ket{\psi}$. 
This leads us to the definition of the \emph{entanglement asymmetry}, given by  
\begin{equation}\label{eq:definition}
    \Delta S_A^{(n)}\equiv S_n(\rho_{A, Q})-S_n(\rho_A),
\end{equation}
which quantifies the extent of symmetry breaking in the system. Here, $S_n(\rho)$ denotes the $n$-th R\'enyi entropy,  $S_n(\rho) = (1-n)^{-1} \log \mathrm{Tr}(\rho^n)$, as introduced in \cite{Ares:2022koq}. The analytic continuation $n\to 1$ recovers the von Neumann asymmetry~\cite{Calabrese:2009qy}.  
Importantly, Eq.~\eqref{eq:definition} serves as a measure of asymmetry~\cite{chitambar2019quantumresourcetheories} since it is \emph{faithful}, meaning that the entanglement asymmetry is always non-negative $\Delta S_A^{(n)}\geq 0$, and it vanishes if and only if the state is symmetric $[\rho_A, Q_A]=0$. 
Operationally, the symmetrized density matrix is expressible via Fourier transform as  
\begin{equation}
    \rho_{A, Q}=\int_{-\pi}^\pi \frac{{\rm d}\alpha}{2\pi} e^{-i\alpha Q_A} \rho_A e^{i\alpha Q_A}.
\end{equation}
Consequently, its moments are given by  
\begin{equation}\label{eq:FT}
    \mathrm{Tr}(\rho_{A, Q}^n)=\int_{-\pi}^\pi \frac{{\rm d}\alpha_1\cdots{\rm d}\alpha_n}{(2\pi)^n} Z_n(\boldsymbol{\alpha}),
\end{equation}
where $\boldsymbol{\alpha}=\{\alpha_1,\dots,\alpha_n\}$ and $Z_n(\boldsymbol{\alpha})$, known as \emph{charged moments}, are defined as  
\begin{equation}\label{eq:Znalpha}
    Z_n(\boldsymbol{\alpha})=
    \mathrm{Tr}\left[\prod_{j=1}^n\rho_A e^{i\alpha_{j,j+1}Q_A}\right],
\end{equation}
with $\alpha_{ij} \equiv \alpha_i - \alpha_j$ and $\alpha_{n+1}=\alpha_1$.  
As we discuss below, computing $Z_n(\boldsymbol{\alpha})$ is significantly more tractable than directly evaluating Eq.~\eqref{eq:definition}.

\subsection{Quantum Mpemba effect }
The most striking feature this quantity can detect is the so-called \emph{quantum Mpemba effect}. 
As mentioned in the introduction, this phenomenon describes the counterintuitive situation where greater symmetry breaking at the initial time $t=0$ leads to a faster symmetry restoration.
Specifically, given the entanglement asymmetry for two states, $\rho_1$ and $\rho_2$, the Mpemba effect occurs when $\Delta S_A^{(n)}(\rho_1,t=0) > \Delta S_A^{(n)}(\rho_2,t=0)$, yet at late time $\Delta S_A^{(n)}(\rho_1,t\gg1) < \Delta S_A^{(n)}(\rho_2,t\gg1)$. 
This implies the existence of a characteristic time $t_M$ at which the entanglement asymmetry curves for $\rho_1$ and $\rho_2$ cross. This crossing serves as the defining signature of the Mpemba effect.  

Crucially, the eventual symmetry restoration, i.e.,  $\Delta S_A^{(n)}(t\to \infty) = 0$, does not necessarily imply the presence of Mpemba phenomenology. 
By leveraging the quasiparticle picture of entanglement~\cite{calabrese2005evolution,fagotti08,alba2017entanglement,calabrese2018entanglement}, Refs.~\cite{Rylands:2023yzx,Murciano:2023qrv} have established criteria to predict \textit{a priori} whether the Mpemba effect occurs in integrable systems. These criteria depend on the charge probability distribution of the initial state and, in free systems, can be reformulated in terms of Cooper pairs, which drive the breaking of particle-number symmetry.
Beyond integrable systems, a condition for the quantum Mpemba effect has also been identified in the presence of weak dephasing or dissipative processes for different initial states~\cite{Caceffo:2024jbc,Ares:2024nkh}. In these cases, the relevant criterion depends on the slowest Cooper pairs that leave the subsystem $A$.   
When the dynamics are governed by chaotic quandum dynamics, the system lacks a quasiparticle description. Nevertheless, the relationship between asymmetry and operator spreading provides an alternative criterion to explain the occurrence of the Mpemba effect, applicable to non-integrable systems~\cite{Turkeshi:2024juo}. 

%It is worth mentioning another setup where the symmetry is restored, starting from a state that is invariant under $U(1)$, and then it evolves with random unitary dynamics that do not preserve it. 

\subsection{Lack of symmetry restoration}
A straightforward scenario in which symmetry is not restored arises when the system evolves under dynamics that do not preserve it. For instance, Refs.~\cite{Ares:2023ggj,Ares:2025ljj,yu2025symmetrybreakingdynamicsquantum} examine the evolution of a symmetry that is initially conserved but later broken by random unitary dynamics. Their findings indicate that when the subsystem is sufficiently large relative to the entire system, the asymmetry converges to a nonzero value, signaling the absence of symmetry restoration. Conversely, for smaller subsystems, an emergent $U(1)$ symmetry restoration occurs as the system relaxes to a maximally mixed symmetric state.  
A less trivial case where symmetry is not restored occurs in a quantum quench starting from a tilted antiferromagnetic (N\'eel) state. Despite the evolution being governed by a $U(1)$-symmetric Hamiltonian, the $U(1)$ symmetry is not necessarily restored at late times. In integrable quenches, this behavior can be attributed to the activation of a non-Abelian set of conserved charges, all of which break the $U(1)$ symmetry. As a result, the entanglement asymmetry saturates to a nonzero constant value~\cite{Ares:2023kcz}. A similar effect is observed when generic unitary evolution is implemented via random circuits with a global $U(1)$ symmetry~\cite{Turkeshi:2024juo,liu2024symmetry}.  
Having outlined the key features of the entanglement asymmetry evolution and its connection to symmetry breaking, we now turn to an analysis of how these aspects are influenced by non-Hermitian dynamics.

\section{Results}
\label{sec:results}
\subsection{Non-Hermitian $XY$ chain}\label{sec:prot1}
Consider the ground state $\ket{\psi(0)}$ of the $XY$ spin chain of length $L$, that in fermionic variables reads
% The first non-Hermitian protocol that we analyze is the following. We start from the ground state $\ket{\psi(0)}$ of the $XY$ spin chain of length $L$ that, for later convenience, we write down directly in terms of fermionic variables, 
\begin{equation}
\label{eq:Ham_prot1}
H(\kappa,h)=-\sum_{j=1}^{L}\left(c^{\dagger}_j c_{j+1}+\kappa c^{\dagger}_jc^{\dagger}_{j+1} +\mathrm{h.c.}-2h c^{\dagger}_jc_j\right)\;,
\end{equation}
with anti-periodic boundary conditions $c_{L+1}=-c_{1}$ imposed. 
For $t>0$ the system is subject to continuous monitoring of the particle density $n_j=c^{\dagger}_jc_j$ with rate $\gamma$, such that the evolution of the system follows the
stochastic Schr\"odinger equation ~\cite{turkeshi2023entanglementandcorrelation}
\begin{equation}
\label{eq:nonHermitianSchroe}
    \frac{d\ket{\psi(t)}}{dt}=-iH(0,0) dt\ket{\psi(t)}-\frac{\gamma}{2}dt\sum_i(n_i-\braket{n_i(t)})\ket{\psi(t)}+\sum_id\mathcal{N}_i\left(\frac{n_i}{\sqrt{\braket{n_i(t)}}}-1\right)\ket{\psi(t)}\;,
\end{equation}
where $d\mathcal{N}_i=0,1$. As we anticipated in the introduction, the
stochastic Schr\"odinger equation marks the different mathematical setup of previous studies on the quantum Mpmeba effect in open systems, whose dissipative dynamics is described by a Lindblad equation. 
Throughout this work we study the post-selected dynamics $d\mathcal{N}_i=0$ at any time $t$, the so-called \textit{no-click limit}.
In this case, Eq.~\eqref{eq:nonHermitianSchroe} simplifies considerably, with the dynamics fully captured by the non-Hermitian Hamiltonian
\begin{equation}
\label{eq:evolution Ham}
    H_{\mathrm{ev}}=H(0,0)-i\frac{\gamma}{2}\sum_{j=1}^L c^{\dagger}_jc_j,
\end{equation}
and given by
\begin{equation}\label{eq:state}
    \ket{\psi(t)}=\frac{e^{-iH_{\mathrm{ev}}t}\ket{\psi(0)}}{||e^{-iH_{\mathrm{ev}}t}\ket{\psi(0)}||}\;.
\end{equation} 
To understand the physical meaning of the parameters in this protocol, we first note that the anisotropy $\kappa$ at $t=0$ is responsible for breaking the  $U(1) $ symmetry. Meanwhile, the non-Hermitian coupling $\gamma$ not only sets the measurement frequency but also acts as a dissipation rate~\cite{turkeshi2023entanglementandcorrelation}.
When $\ket{\psi(0)}$ is a Gaussian state, Eq.~\eqref{eq:state} describes a Gaussian evolution. As a result, the reduced density matrix $\rho_A(t) = \mathrm{Tr}_B \ket{\psi(t)}\bra{\psi(t)}$ for a subsystem $A$ of size $\ell$ is determined by the time-dependent two-point correlation matrix restricted on $A$. 
This correlation matrix is the central object to computing both the entanglement and the asymmetry, making its explicit derivation the focus of the next section.
While the following formulas could, in principle, be extracted from the implicit expressions in \cite{turkeshi2023entanglementandcorrelation}, we provide a detailed derivation here. This step-by-step approach will also prove useful in obtaining explicit correlators for the second protocol studied in this work.

\subsubsection{Correlation functions}
\label{subsec:correlation_1stquench}
The entanglement entropies and the entanglement asymmetry in the non-Hermitian quantum quench protocol~\eqref{eq:state} are fully determined by the two-point correlation matrix, expressed in terms of the fermionic operators $\boldsymbol{c}_j=(c_j^\dagger, c_j)$ 
\begin{equation}\label{eq:corr}
\Gamma_{jj'}=2\mathrm{Tr}\left[\rho_A\boldsymbol{c}_j^\dagger
 \boldsymbol{c}_{j'}\right]-\delta_{jj'}
 =
 \begin{pmatrix}
      2 \langle c_j
 c^\dagger_{j'}\rangle- \delta_{jj'}  && 2\langle c_j
 c_{j'}\rangle
        \\
    2 \langle c^\dagger_j
 c^\dagger_{j'}\rangle && 2 \langle c^\dagger_j
 c_{j'}\rangle- \delta_{jj'}
    \end{pmatrix}
 ,
\end{equation}
where the expectation value is defined as $\langle \cdot\rangle=\mathrm{Tr}\left[\rho_A\cdot\right]$. 
To compute $\Gamma_{jj'}$, we first perform a Fourier transformation, diagonalize the Hamiltonian, and then solve the equations of motion for the fermionic modes, cf.~\cite{turkeshi2023entanglementandcorrelation}. 
We now outline this derivation, starting from a finite chain with an even number of sites $L$. 
Consistently with the anti-periodic boundary conditions in~\eqref{eq:Ham_prot1}, the momenta are chosen as 
\begin{equation}
\label{eq:momenta_ABC}
    k=\pm\frac{(2n-1)\pi}{L}\,,
    \quad n=1,2,\dots,\frac{L}{2}.
\end{equation}
By Fourier transforming the fermionic variables,
we can rewrite the Hamiltonian as 
\begin{equation}
    H(\kappa,h)-i\frac{\gamma}{2}\sum_jc^{\dagger}_jc_j=\sum_{k>0} \left[
    \lambda_k\left( c_{-k}c^\dagger_{-k} 
     - c^\dagger_kc_k
     \right)
     +i\theta_k
     \left( 
     c_{-k}c_{k}
    -c^\dagger_kc^\dagger_{-k}  
    \right)
    \right]\equiv \sum_{k>0}H_k\,,\quad  c_j=\frac{1}{\sqrt{L}}\sum_k e^{ikj} c_k,
\label{eq:Hamiltonian_momentum}
\end{equation}
with $\lambda_k=2\cos k-2h+i\gamma/2 \,,$
and $ \theta_k=2\kappa\sin k\,$.
The first step in solving the quench is to determine the ground state of Eq.~\eqref{eq:Hamiltonian_momentum} with $\gamma=0$, which serves as the initial state at $t=0$. 
Since $H(\kappa,h)$ is a sum over the positive momenta, the ground state decomposes in momentum space as 
\begin{equation}
\label{eq:factorized_GS}
\vert \psi(0)\rangle=\bigotimes_{k>0} \vert \psi_k(0)\rangle\,.
\end{equation}
Furthermore, each $H_k$ acts non-trivially only within a two-dimensional subspace of the fixed-momentum Hilbert space, spanned by $\vert 0\rangle$ and $c^\dagger_{-k} c^\dagger_k\vert 0\rangle$. Consequently, $\vert \psi_k(0)\rangle$ must take the form
\begin{equation}\label{eq:prot1k}
\vert \psi_k(0)\rangle=
\left(u_k(0)+v_k(0)c^\dagger_{-k} c^\dagger_k\right)\vert 0\rangle \,,
\end{equation}
for certain $u_k(0)$ and $v_k(0)$ determined by the ground state $\ket{\psi(0)}$. 
Concretely, we note that the action of $H_k$ on $\vert \psi_k(0)\rangle$ is equivalent to the action of a $2\times 2$ matrix on a two-dimensional vector
\begin{equation}
\label{eq:2drep_Hamiltonian}
 H_k \vert \psi_k(0)\rangle\quad \Leftrightarrow \quad  
 \mathcal{M}_k(\kappa, h,0) \begin{pmatrix}
         u_k(0)
        \\
      v_k(0)
    \end{pmatrix}\,,
\qquad
\mathcal{M}_k(\kappa, h,\gamma)=\begin{pmatrix}
       \lambda_k  & -i \theta_k
        \\
     i \theta_k & -\lambda_k
    \end{pmatrix}.
\end{equation}
In summary, for any $k$, the coefficients $u_k(0)$ and $v_k(0)$ are the components of the normalized eigenvector of $\mathcal{M}_k$ with the smallest eigenvalue, explicitly
\begin{equation}
\label{eq:uk0}
    u_k(0)=i{\rm sgn}  (k)\sqrt{\frac{h-\cos k+\sqrt{(h-\cos k)^2+\kappa^2\sin^2 (k)}}{2\sqrt{(h-\cos k)^2+\kappa^2\sin^2 (k)}}}\,,
\end{equation}
\begin{equation}
\label{eq:vk0}
    v_k(0)=\sqrt{\frac{\cos k-h+\sqrt{(h-\cos k)^2+\kappa^2\sin^2 (k)}}{2\sqrt{(h-\cos k)^2+\kappa^2\sin^2 (k)}}}\,.
\end{equation}
The above Eqs.~\eqref{eq:uk0} and~\eqref{eq:vk0} completely determine the ground state of $H(\kappa,h)$. 
We now determine the evolution of this ground state through the Hamiltonian~\eqref{eq:state}.
Since the evolution Hamiltonian retains the same structure as Eq.~\eqref{eq:Hamiltonian_momentum} in the momentum space, the evolved state at time $ t $ maintains the factorized form
\begin{equation}
\label{eq:timeevolved_state}
\vert \psi(t)\rangle=\bigotimes_{k>0} \vert \psi_k(t)\rangle
=\bigotimes_{k>0}
\left(\frac{u_k(t)+v_k(t)c^\dagger_{-k} c^\dagger_k}{\sqrt{\vert u_k(t)\vert^2+\vert v_k(t)\vert^2}}\right)\vert 0\rangle 
\,,
\end{equation}
where the denominator ensures proper normalization.
% We note that, unlike in the purely unitary case, the dynamics governed by the non-Hermitian Hamiltonian~\eqref{eq:evolution Ham} necessitates the inclusion of measurement back-action to preserve the norm, explicitly enforced by the denominator in Eq.~\eqref{eq:state}. 
To determine the time evolution of the coefficients  $u_k(t) $ and  $v_k(t)$, we express and solve the Schrödinger equation in the two-dimensional representation of Eq.~\eqref{eq:2drep_Hamiltonian}, which takes the form
\begin{equation}
 i\begin{pmatrix}
         \dot u_k(t)
        \\
      \dot v_k(t)
    \end{pmatrix}= \mathcal{M}_k(0, 0,\gamma) \begin{pmatrix}
         u_k(t)
        \\
      v_k(t)
    \end{pmatrix} \,.
\end{equation}
When $h=\kappa=0$, $\mathcal{M}_k$ is diagonal and the equation is straightforward to solve, leading to
\begin{equation}
\label{eq:ukt}
  u_k(t)= e^{-2{\rm i}t \cos k+\gamma t/2} u_k(0)\,,\quad
   v_k(t)=  e^{2{\rm i}t \cos k-\gamma t/2} v_k(0)  \,,
\end{equation}
where the initial conditions $u_k(0)$ and $v_k(0)$ are given by Eqs.~\eqref{eq:uk0} and \eqref{eq:vk0} respectively.

Having determined both the ground state and its evolution, we are now in position to compute the entries of the two-point correlation matrix in Eq. \eqref{eq:corr} for the quench under consideration. Using the form \eqref{eq:timeevolved_state} and taking the thermodynamic limit $L\to \infty$, we obtain
\begin{equation}
\label{eq:Gammawithsymbol}
    \Gamma_{ll'}(t)=\int_{-\pi}^\pi \frac{dk}{2\pi}e^{-ik(l-l')}
    \mathcal{G}(k,t),
\end{equation}
where
\begin{equation}\label{eq:symbol}
    \mathcal{G}(k, t)=\left(\begin{array}{cc} 
    n(k, t) & g(k,t) \\
    g^*(k,t) & -n(k, t) \\
    \end{array}\right),
\end{equation}
with
\begin{equation}
    n(k,t)=\frac{\vert u_k(t)\vert^2-\vert v_k(t)\vert^2}{\vert u_k(t)\vert^2+\vert v_k(t)\vert^2},\qquad
g(k,t)=\frac{-2 u_k^*(t)v_k(t)}{\vert u_k(t)\vert^2+\vert v_k(t)\vert^2}.
\end{equation}
Crucially, the correlation matrix in Eq.~\eqref{eq:Gammawithsymbol} exhibits a block Toeplitz structure with $2\times 2$ blocks. The $2\times 2$ matrix $\mathcal{G}(k, t)$ is referred to as symbol and, using Eqs. \eqref{eq:uk0}, \eqref{eq:vk0}, \eqref{eq:ukt}, its entries can be rewritten in terms of the quench parameters $\kappa$, $h$ and $\gamma$ as
\begin{equation}
\label{eq:fg_def}
\begin{split}
    n(k,t)=\frac{\sinh(\gamma t)\sqrt{(\cos (k)-h)^2+\kappa ^2 \sin ^2(k)}-\cosh(\gamma t)(\cos k-h)}{\cosh(\gamma t)\sqrt{(\cos (k)-h)^2+\kappa ^2 \sin ^2(k)}-\sinh(\gamma t)(\cos k-h)}&,\\
g(k,t)=\frac{i\kappa e^{4 i t \cos k}\sin k}{\cosh(\gamma t)\sqrt{(\cos (k)-h)^2+\kappa ^2 \sin ^2(k)}-\sinh(\gamma t)(\cos k-h)}&.
\end{split}
\end{equation}
In the limit $\gamma\to 0$, we recover the correlator for a quench to the  $XX $ Hamiltonian, as derived in Ref.~\cite{fagotti08}.
The charge moments of Eq.~\eqref{eq:Znalpha} are expressed in terms of the two-point correlation matrix $\Gamma$ by leveraging the properties of Gaussian operators \cite{Fagotti2010}
\begin{equation}\label{eq:numerics} Z_n(\boldsymbol{\alpha})=\sqrt{\det\left[\left(\frac{I-\Gamma}{2}\right)^n
  \left(I+\prod_{j=1}^n W_j\right)\right]},
\end{equation}
where $W_j=(I+\Gamma)(I-\Gamma)^{-1}e^{i\alpha_{j,j+1} n_A}$ and $n_A$ is a diagonal matrix with $(n_A)_{2j,2j}=1$, while $(n_A)_{2j-1,2j-1}=-1$ for $j=1, \cdots, \ell$. 
To compute the entanglement asymmetry, we first evaluate the charged moments  $Z_n(\boldsymbol{\alpha})$ using Eq.~\eqref{eq:numerics}. Then, substituting this result into Eq.~\eqref{eq:FT}, we numerically compute   $\Delta S_A^{(n)}$.

\textbf{Early time behavior.---}
At $ t = 0 $, we recover the previously established results for the entanglement asymmetry found in Ref.~\cite{Murciano:2023qrv}, which we include here for completeness.
For a subsystem of large size $ \ell\to\infty $, the asymmetry is given by
\begin{equation}\label{eq:Deltaspn}
\Delta S_A^{(n)}=\frac{1}{2}\log \ell+\frac{1}{2} \log \frac{\pi s(\kappa,h)n^{1/(n-1)}}{4}+O(\ell^{-1}),
\end{equation}
where 
\begin{equation}\label{eq:g_gs_explicity}
 s(\kappa, h)=\begin{cases}
               \frac{\kappa}{\gamma+1},\quad &|h|\leq1, \\
               \frac{\kappa^2}{1-\kappa^2}\left(\frac{|h|}{\sqrt{h^2+\kappa^2-1}}-1\right), \quad& |h|>1.
              \end{cases}
\end{equation}
We remark that if we focus along the line $h^2+\kappa^2=1$, the ground state of Eq. \eqref{eq:Ham_prot1} is the tilted ferromagnetic state with tilting angle $\cos(2\theta)=(1-\kappa)/(1+\kappa)$. As Refs. \cite{Murciano:2023qrv,Ares:2022koq,Rylands:2023yzx} have shown, for the tilted ferromagnetic state the entanglement asymmetry grows monotonically as a function of $\theta$ for $\theta \in [0,\pi/2]$. This result is quite intuitive because larger values of $\theta$ correspond to larger symmetry breaking.

\textbf{Late time behavior.---}
\begin{figure}[t!]
\hspace{-0.0cm}
    \includegraphics[width=0.54\textwidth]{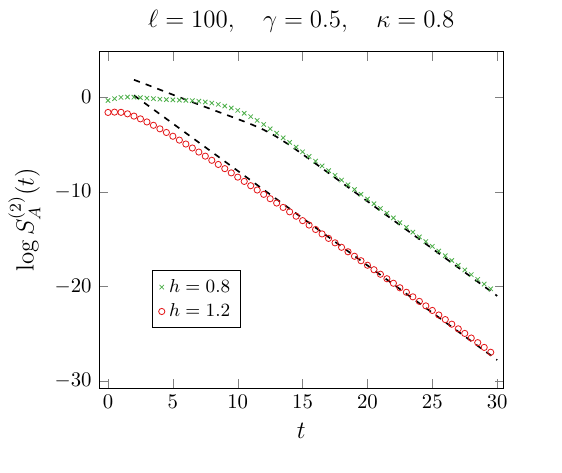}
     \includegraphics[width=0.54\textwidth]{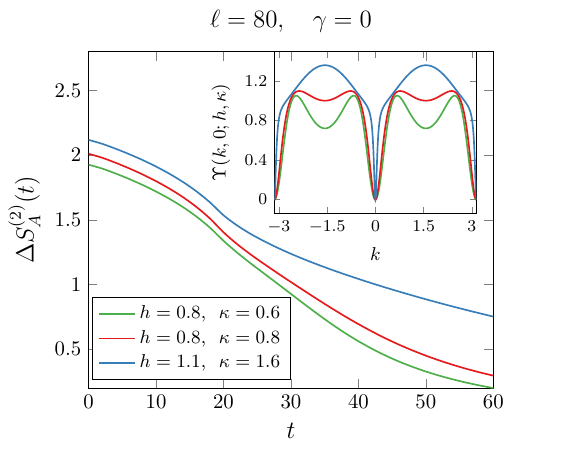}
      \includegraphics[width=0.54\textwidth]{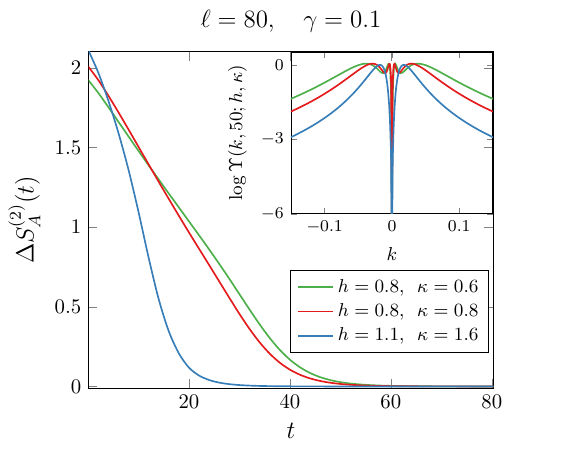}
     \includegraphics[width=0.54\textwidth]{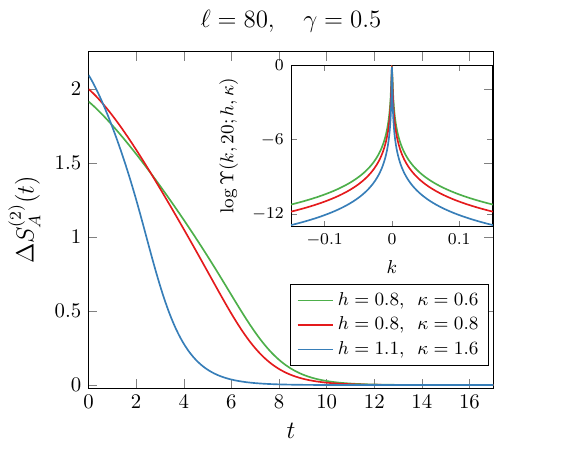}
    \caption{Dynamics of the second Rényi entropy (top left panel) and second Rényi asymmetry (top right and bottom panels) after a quench from the ground state of the XY chain. The exponential decay of the second Rényi entropy in the top left panel occurs with a rate equal to $\gamma$ (dashed lines).
    The second Rényi asymmetry is shown for different values of $\gamma$ and three choices of the parameters $h$ and $\kappa$ in each of the remaining panels. In the case of unitary dynamics (top right panel), we do not observe the quantum Mpemba effect. On the other hand, when the non-Hermitian term in the evolution Hamiltonian is turned on, the quantum Mpemba effect occurs. This phenomenon can be understood by looking at the density of Cooper pairs reported in the insets.
   }
    \label{fig:1stquench}
\end{figure}
Since the evolution is now non-Hermitian, the quench no longer admits a simple quasiparticle description. 
However, we can still extract meaningful physical insights by examining the late-time regime. To do so, we analyze the entanglement dynamics along the time evolution \eqref{eq:state}.
As $t\to \infty$, the terms $e^{\pm2 i t \cos k}$ in Eq. \eqref{eq:ukt} average to zero and only the diagonal components of the symbol \eqref{eq:Gammawithsymbol} do not vanish. 
Leveraging this fact, we compute the late-time behavior of the R\'enyi entropy in the limit of large subsystem size $ \ell\to\infty$
\begin{equation}\label{eq:latetime}
    S_n(\rho_A,t\to \infty)=\frac{\ell}{2(1-n)}\int_{-\pi}^{\pi} \frac{dk}{2\pi}\log \left[\left(\frac{1+|n(k,t)|}{2}\right)^n+\left(\frac{1-|n(k,t)|}{2}\right)^n\right],
\end{equation}
where $ n(k,t)$, the diagonal entry of the symbol in Eq.~\eqref{eq:symbol}, represents the density of occupied modes with momentum $k$. 
Eq.~\eqref{eq:latetime} reveals an entanglement evolution that differs significantly from the one reported in Ref.~\cite{turkeshi2023entanglementandcorrelation}.
There, depending on the measurement rate $\gamma$, the late time entanglement entropy was found to saturate to either a stationary value that scales logarithmically with subsystem size or a value independent of $\ell$. 
The key distinction here is the choice of the initial state: instead of a simple ferromagnetic state, we start from the ground state of Eq.~\eqref{eq:Ham_prot1}. This difference crucially affect the entanglement dynamics. Specifically, after an initial growth, the entanglement entropy undergoes an exponential decay over time, with a decay rate set by $\gamma$. 
We present this behavior for the second R\'enyi entropy in the top-left panel of  Fig.~\ref{fig:1stquench}, where the dashed line represent an exponential decay with rate $\gamma$. 
An intuitive explanation for this phenomenon arises in the regime  $\gamma \gg 1 $: in this limit, the overlap between the stationary state (where all spins align along the $z$-direction) and the initial state is exponentially suppressed with system size. As a result, in the thermodynamic limit considered here, the overlap ultimately vanishes. This contrasts with the scenario in Ref.~\cite{turkeshi2023entanglementandcorrelation} where the quench starts from a ferromagnetic state, and the overlap remains nonzero, leading to a qualitatively different entanglement evolution.

We now shift our focus to the behavior of the entanglement asymmetry. 
Without a clear quasiparticle picture of entanglement spreading, we can only verify that, at late time, the symmetry is restored for any parameter $h,\kappa,\gamma$, since the asymmetry is fully determined by the exponentially decaying $S_n(\rho_A,t\to \infty)$, cf. Eq.~\eqref{eq:latetime}. 
Still, the fermionic evolution allows to numerically simulate the behavior of $\Delta S_A^{(n)}$ using Eqs. \eqref{eq:numerics} and \eqref{eq:FT}. 
We report the results in Fig.~\ref{fig:1stquench}. We note that the asymmetry decays exponentially at a rate $2\gamma$, similar to what happens in systems with balanced gain and loss dissipation \cite{Ares:2024nkh}. 

However, an unexpected phenomenon emerges in this case. For any measurement rate $\gamma>0$, the bottom panels in Fig.~\ref{fig:1stquench} clearly show that for fixed $\gamma$, the curves for different initial parameters $(h,\kappa)$ cross. 
This suggests that even a small  $\gamma > 0 $ induces a quantum Mpemba effect—a phenomenon absent in systems with balanced gain and loss, regardless of dissipation strength \cite{Ares:2024nkh}.
Although we lack an analytical prediction for the late-time behavior of the asymmetry, we conjecture that it remains closely related to the density of Cooper pairs with momentum $ k $, given by $|\braket{c^{\dagger}_k c^{\dagger}_{-k}}|,$ which are responsible for symmetry breaking in the particle number sector. 
This interpretation is consistent with cases where a quasiparticle picture is available~\cite{Murciano:2023qrv}.
In the notation used in this section, the Cooper pair density is encoded in the function $g(k,t)$ defined in \eqref{eq:fg_def}. For clarity, we now explicitly denote its dependence on the initial state parameters, writing it as $g(k,t;h,\kappa)$.
According to the criterion outlined above, the quantum Mpemba effect occurs when the state that initially breaks less the symmetry contains a smaller number of Cooper pairs, but it has a larger density of Cooper pairs at long times around the modes with the slowest velocity $v(k)=-\sin(k)$, which are $k=0,\pi$. 
In formulas, this implies that at large times
\begin{equation}\label{eq:crit}
    \Delta S_A^{(n)}(\rho_1,t)<\Delta S_A^{(n)}(\rho_2,t) \Leftrightarrow |g(k,t;h_1,\kappa_1)|^2 < |g(k,t;h_2,\kappa_2)|^2,
\end{equation}
for $k$ close to $0 $ and $\pi$.
Since these Cooper pairs are the ones responsible for the symmetry breaking also in our protocol, after a change of variable $k\to k-\pi$, we plot $\Upsilon(k,t;h,\kappa)\equiv|g(k,t;h,\kappa)|^2+|g(k,t;-h,\kappa)|^2$ in the insets of Fig.~\ref{fig:1stquench}. Indeed, the transformation $h\to -h$ on $g$ in \eqref{eq:fg_def} is equivalent to $k\to k-\pi$. 
The late-time behavior in Fig.\,\ref{fig:1stquench} confirms that the condition in Eq.~\eqref{eq:crit} is satisfied, thereby explaining the presence, or absence, of the quantum Mpemba effect in the corresponding panels. 
We conclude by emphasizing that a nonzero measurement rate $\gamma \neq 0$  can induce a quantum Mpemba effect, even in regimes where it has never been observed before. We dub this phenomenon as a \textit{genuine} measurement-induced Mpemba effect.

\subsection{Anisotropic SSH chain}
\label{sec:2ndprotocol}
The second quench we analyze in this section differs from the previous one in a crucial way: it involves a model where translational symmetry is reduced to two-site translations. While this might seem like a minor modification, it has a significant impact on symmetry restoration. 
To explore this effect, we consider the SSH model, which inherently exhibits two-site translational symmetry. Additionally, we explicitly break the global $ U(1) $ symmetry by introducing a pairing term. The initial state for this quench is chosen as the ground state of this model, whose Hamiltonian is
\begin{align}\label{eq:Ham2}
H(h,\kappa)&=-(1+h/2)\sum_{j=1}^{L/2}[c^{\dagger}_{2j-1}c_{2j}]   - (1-h/2)\sum_{j=1}^{L/2-1}[c^{\dagger}_{2j}c_{2j+1}]-\kappa \sum_{j=2}^{L} [c^{\dagger}_{j-1}c^{\dagger}_{j}]\\ &\quad  + (1-h/2) c^{\dagger}_{L}c_{1} + \kappa c^{\dagger}_{L}c^{\dagger}_{1} + \text{h.c.},
\nonumber
\end{align}
which breaks the  $U(1) $ symmetry for  $\kappa \neq 0 $ and one-site translational invariance for $ h \neq 0 $.
Also for this protocol, we consider a continuously monitored system, whose no-click limit follows Eq.~\eqref{eq:state} with the non-Hermitian Hamiltonian~\cite{legal2023volume}
\begin{equation}
    \label{eq:Ham2_ev}
H_{\rm ev}=H(h_{\rm ev},0)+i\frac{\gamma}{2}\sum_j[c^{\dagger}_{2j}c_{2j}-c^{\dagger}_{2j+1}c_{2j+1}].
\end{equation}
% Thus, the evolved state takes the same form of Eq. \eqref{eq:state}. 
We immediately observe that the Hamiltonians remain quadratic in the fermionic operators. Consequently, we can apply techniques for Gaussian states similar to those used in Sec.~\ref{sec:prot1}.

\subsubsection{Correlation functions}
We start the analysis by considering the behavior of the state at time $t=0$, i.e. we look for the ground state of \eqref{eq:Ham2}. One can diagonalize this Hamiltonian, by first introducing the operators
\begin{equation}
\label{eq:oddeven_FT}
c_{2j-1}=\frac{1}{\sqrt{L}}\sum_k e^{i k j}  c_{k,o}, \quad c_{2j}=\frac{1}{\sqrt{L}}\sum_k e^{i k j}  c_{k,e},
\end{equation} 
where the set of momenta is given by Eq. \eqref{eq:momenta_ABC}. Plugging Eq. \eqref{eq:oddeven_FT} into the Hamiltonian \eqref{eq:Ham2}, we find
\begin{equation}
\label{eq:Ham2_momentum}
H(h,\kappa)=\sum_{k>0}  \left(c^{\dagger}_{k,o} c^{\dagger}_{k,e} c_{-k,o} c_{-k,e}  \right)\mathcal{H}_k(h,\kappa) \left(\begin{array}{c} 
    c_{k,o}  \\
   c_{k,e} \\
  c^{\dagger}_{-k,o} \\
 c^{\dagger}_{-k,e}  \\
    \end{array}\right)
    \equiv
   \sum_{k>0}  H_k(h,\kappa),
\end{equation}
where $H_k(h,\kappa)$ is the single-particle Hamiltonian in momentum space
\begin{equation}
\mathcal{H}_k(h,\kappa)=\begin{pmatrix}
    0 & A_{-k}(h) &0& B_{-k}(\kappa) \\
    A_{k}(h) & 0 & -B_{k}(\kappa) & 0 \\
   0 &  -B_{-k}(\kappa) & 0& -A_{-k}(h) \\
   B_{k}(\kappa) &0 & -A_{k}(h)&0 \\
    \end{pmatrix} ,
\end{equation}
and it is implicitly defined in terms of the functions
\begin{equation}
\label{eq:AandB_def}
   A_{k}(h) \equiv-\left(1-\frac{h}{2}\right)e^{ik}-\left(1+\frac{h}{2}\right)\,,\quad 
   B_{k}(\kappa) \equiv \kappa\left(e^{ik}-1\right)\,.
\end{equation}
The structure of Eq. \eqref{eq:Ham2_momentum} suggests that the ground state has the form of the state \eqref{eq:factorized_GS}. Studying how $H_k(h,\kappa)$ in Eq. \eqref{eq:Ham2_momentum} acts on the independent combinations of the fermionic bilinears, 
we deduce that the general form of  $\vert\psi_k(0)\rangle$ is
\begin{equation}
\label{eq:initialstate_k_2ndquench}
\begin{split}
\vert\psi_k(0)\rangle
=&\left(u_{1,k}(0)+u_{2,k}(0)c^\dagger_{k,o} c^\dagger_{-k,o}+u_{3,k}(0)c^\dagger_{k,e} c^\dagger_{-k,e}
\right.\\
&
\left.
+ u_{4,k}(0)c^\dagger_{k,o} c^\dagger_{-k,o}c^\dagger_{k,e} c^\dagger_{-k,e}+u_{5,k}(0)c^\dagger_{k,o} c^\dagger_{-k,e}+u_{6,k}(0)c^\dagger_{k,e} c^\dagger_{-k,o}\right)\vert 0\rangle\,.
 \end{split}
\end{equation}
Also in this case, we can explicitly verify that there are no other operators arising in $H_k(h,\kappa)\vert\psi_k(0)\rangle$. 
We stress that the main difference of $\vert\psi_k(0)\rangle$ compared to Eq. \eqref{eq:prot1k} is due to the fact that now $H_k(h,\kappa)$ acts non-trivially on all the possible even-odd combinations of bilinear terms, and also on the quartic term. For this reason, the procedure exploited in Sec.\,\ref{subsec:correlation_1stquench} to determine the unknown coefficients has to be adapted. We discuss this generalization in Appendix \ref{app:Details}, where we also detail the explicit expressions of $u_{j,k}(0)$.

At time $t>0$, the post-quench state also admits a factorized structure \eqref{eq:factorized_GS}, with
\begin{align}\label{eq:evolvedstate_2ndquench}
\vert\psi_k(t)\rangle
=&\frac{1}{\vert\vert\psi_k(t)\vert\vert^2}\left(u_{1,k}(t)+u_{2,k}(t)c^\dagger_{k,o} c^\dagger_{-k,o}+u_{3,k}(t)c^\dagger_{k,e} c^\dagger_{-k,e}
\right.
\\
&
\left.
+ u_{4,k}(t)c^\dagger_{k,o} c^\dagger_{-k,o}c^\dagger_{k,e} c^\dagger_{-k,e}+u_{5,k}(t)c^\dagger_{k,o} c^\dagger_{-k,e}+u_{6,k}(t)c^\dagger_{k,e} c^\dagger_{-k,o}\right)\vert 0\rangle\,,
\nonumber
\end{align}
where
$\vert\vert\psi_k(t)\vert\vert^2$ ensures the normalization at any time step, cf. Appendix \ref{app:Details} for explicit expressions. % The functions of time $u_{j,k}(t)$ can be determined as shown in Appendix \ref{app:Details} generalizing the computation of Sec.\,\ref{subsec:correlation_1stquench}.

\textit{Mutatis mutandis}, from Eq.~\eqref{eq:evolvedstate_2ndquench} we can determine the correlation functions and, consequently, the entanglement and asymmetry properties. 
Due to the two-unit cell translational symmetry of the Hamiltonian~\eqref{eq:Ham2}, it is convenient to arrange the entries of the correlation matrix into blocks of the form
\begin{align}\label{eq:corr}
\Gamma_{jj'}=&2\mathrm{Tr}\left[\rho_A\boldsymbol{c}_j^\dagger
 \boldsymbol{c}_{j'}\right]-\delta_{jj'},
\end{align}
where $j,j'=1,\dots,L/2$, the even number $L$ is the system size, and we introduced $\boldsymbol{c}_j=(c_{2j-1},c_{2j},c^\dagger_{2j-1},c^\dagger_{2j})$. 
Using the translational invariance with respect to two-sites unit cells and the evolved state \eqref{eq:evolvedstate_2ndquench}, and taking the thermodynamic limit, we find 
\begin{equation}
\label{eq:corr_time t}
   [\Gamma_{\rm d}(t)]_{jj'}=\int_{-\pi}^\pi \frac{dk}{2\pi}e^{-ik(j-j')} \mathcal G_{\rm d}(k,t),
\end{equation}
where the $4\times 4$ symbol reads
\begin{equation}
\label{eq:Symbol_2nd_quench}
    \mathcal G_{\rm d}(k,t)=
    \begin{pmatrix}
       \xi(k,t)   && \varphi(k,t)  && \upsilon(k,t)  && \zeta(k,t) 
        \\
     \varphi^*(k,t) &&  -\xi(k,t)  &&-\zeta(-k,t)  && \upsilon^*(k,t) 
 \\
 \upsilon^*(k,t)  &&-\zeta^*(-k,t)  && -\xi(-k,t)&& -\varphi^*(-k,t)
 \\
  \zeta^*(k,t)&&\upsilon(k,t)   &&-\varphi(-k,t) && \xi(-k,t) 
    \end{pmatrix},
\end{equation}
with
\begin{eqnarray}
\label{eq:def_xi}
 \xi(k,t)&\equiv&
 \frac{-\vert u_{1,k} (t)\vert^2+\vert u_{2,k} (t)\vert^2-\vert u_{3,k} (t)\vert^2+\vert u_{4,k} (t)\vert^2+\vert u_{5,k} (t)\vert^2-\vert u_{6,k} (t)\vert^2}{\vert\vert\psi_k(t)\vert\vert^2}
 \,,
 \\
  \varphi(k,t)&\equiv&
  2\frac{ u_{3,k} (t) u_{5,k}^* (t)+u_{6,k} (t) u_{2,k}^* (t)}{\vert\vert\psi_k(t)\vert\vert^2}
  \,,
  \\
 \upsilon(k,t)&\equiv&
 2\frac{u_{1,k} (t) u_{2,k}^* (t)+u_{3,k} (t) u_{4,k}^* (t)}{\vert\vert\psi_k(t)\vert\vert^2}
  \,,
 \\
 \zeta(k,t)&\equiv&
 2\frac{u_{1,k} (t) u_{5,k}^* (t)-u_{6,k} (t) u_{4,k}^* (t)}{\vert\vert\psi_k(t)\vert\vert^2}
 \,.
 \label{eq:def_zeta}
\end{eqnarray}
Using the evolution of the correlation matrix in Eq.~\eqref{eq:corr_time t}, restricted to a subsystem $ A $ consisting of  $\ell $, even, adjacent sites, we now proceed to analyze the entanglement dynamics following the quench described here in the next subsection.

\subsubsection{Hermitian dynamics and lack of symmetry restoration}

With all these ingredients at our disposal, we can first analyze the dynamics of the unitary quench, where we prepare the system in the ground state of Eq. \eqref{eq:Ham2} and we let it evolve through Eq. \eqref{eq:Ham2_ev} with $\gamma=0$. In this case, to study the entanglement entropies and the entanglement asymmetry, we can resort to the quasiparticle picture. To apply this technology, we need the expressions of the charged moments at large time $t\to\infty$ and at $t=0$.
The subsystem $A$ is an interval of even size $\ell$.
Using Eq. \eqref{eq:numerics} with the correlation matrix \eqref{eq:corr} restricted to $A$ and exploiting the properties of the Toeplitz matrices detailed in Appendix \ref{app:Toeplitz}, in the limit of large subsystem sizes, $\ell\to\infty$, we find that at time $t\to 0$ the charged moments behave as
\begin{equation}
\label{eq:logZnsmalltimes}
    \log Z_n(\boldsymbol{\alpha},t= 0)\sim \frac{\ell}{4}\int_{-\pi}^{\pi}\frac{d k}{2\pi}\log \det \mathcal{M}^{(n)}_{\boldsymbol{\alpha}}(k,t=0)\,,
\end{equation}
where $\mathcal{M}^{(n)}_{\boldsymbol{\alpha}}$ is a $4 \times 4$ matrix obtained from the symbol \eqref{eq:Symbol_2nd_quench} and it reads
\begin{equation}
    \mathcal{M}^{(n)}_{\boldsymbol{\alpha}}(k,t)=\left(\frac{I-\mathcal G_{\rm d}(k,t)}{2}\right)^n\left[I+\prod_{j=1}^n\frac{I+\mathcal G_{\rm d}(k,t)}{I-\mathcal G_{\rm d}(k,t)}e^{i\alpha_{j,j+1}(\sigma_z\otimes I)}\right].
\end{equation}
We stress that, also for this quench protocol, these and the following analytical predictions are valid in the limit of large subsystem size.
For generic replica index $n$, the expression above is quite involved.
However, we can focus on the case $n=2$, which has also been shown to be accessible in experimental settings~\cite{Joshi:2024sup}. 
Indeed, for $n=2$, the determinant in Eq. \eqref{eq:logZnsmalltimes} has a compact expression in terms of the functions $u_{j,k}(0)$ and $\log Z_2(\boldsymbol{\alpha},t= 0)$ reads
\begin{equation}
\label{eq:logZ2smalltimes}
    \log Z_2(\alpha,t= 0)\sim \frac{\ell}{2}\int_{-\pi}^{\pi}\frac{d k}{2\pi}\log \left[1-4\vert u_{1,k}(0) \vert^2\sin^2\alpha\right] \,.
\end{equation}
The expression \eqref{eq:logZ2smalltimes} clearly shows that $Z_2(\alpha=0,t= 0)=1$, as we would expect since the entanglement of the ground state of Eq. \eqref{eq:Ham2} satisfies an area law behavior. 
In the late time limit $t\to\infty$, we have
\begin{equation}
\label{eq:logZn}
    \log Z_n(\boldsymbol{\alpha},t\to\infty)\sim \frac{\ell}{4}\int_{-\pi}^{\pi}\frac{d k}{2\pi}\log \det \mathcal{M}^{(n)}_{\boldsymbol{\alpha}}(k,t\to\infty)\,.
\end{equation}
To compute $\mathcal{M}^{(n)}_{\boldsymbol{\alpha}}(k,t\to\infty)$, we need the value of the symbol at large times. This can be obtained by averaging in time all the entries of $\mathcal G_{\rm d}(k,t)$. The only surviving terms are the ones depending on  $\varphi(k,t) $ and  $\zeta(k,t)$. We call the time averages of these two functions $\varphi_\infty(k) $ and $\zeta_\infty(k) $, respectively.
When $n=2$, the integrand in the right-hand side of Eq. \eqref{eq:logZn} can be written explicitly as
\begin{equation}
\label{eq:logdetM2}
    \log \det \mathcal{M}^{(2)}_{\alpha}(k,t\to\infty)=2\log\big[m^{(\infty)}_{\alpha}(k)/4\big]  \,,
\end{equation}
where
\begin{equation}
\label{eq:m2_def}
m^{(\infty)}_{\alpha}(k)=
1+\vert\zeta_\infty(k) \vert^2 \left(2\cos(2 \alpha)+\vert\zeta_\infty(k) \vert^2\right)+
\vert\varphi_\infty(k) \vert^2 \left(2+\vert\varphi_\infty(k) \vert^2\right)-2 {\rm Re}\left[\left(\varphi^*_\infty(k)\right)^2\zeta^2_\infty(k)\right]
,
\end{equation}
and we have used that $\zeta_{\infty}(-k)=\zeta^*_{\infty}(k)$ and $\varphi_{\infty}(-k)=\varphi^*_{\infty}(k)$. 
Thus, we can rewrite Eq. \eqref{eq:logZn} with $n=2$ as
\begin{equation}
\label{eq:quasiptclZ_2}
\log Z_2(\alpha,t\to\infty)\sim \frac{\ell}{2}\int_{-\pi}^{\pi}\frac{d k}{2\pi}\log\left( m^{(\infty)}_{0}(k)/4-\vert\zeta_\infty(k)\vert^2\sin^2\alpha\right).
\end{equation}

\begin{figure}[t!]
\hspace{-0.6cm}
    \includegraphics[width=0.54\textwidth]{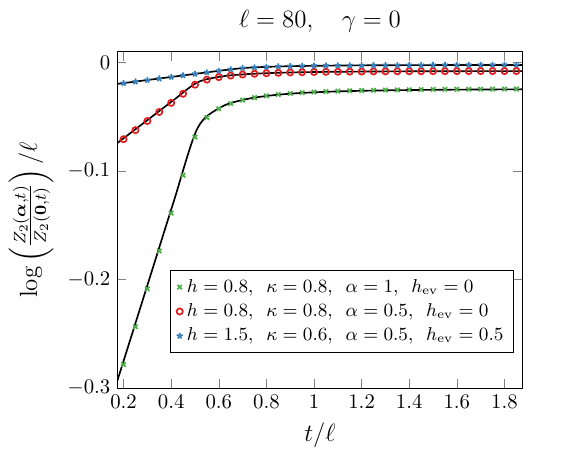}
     \includegraphics[width=0.54\textwidth]{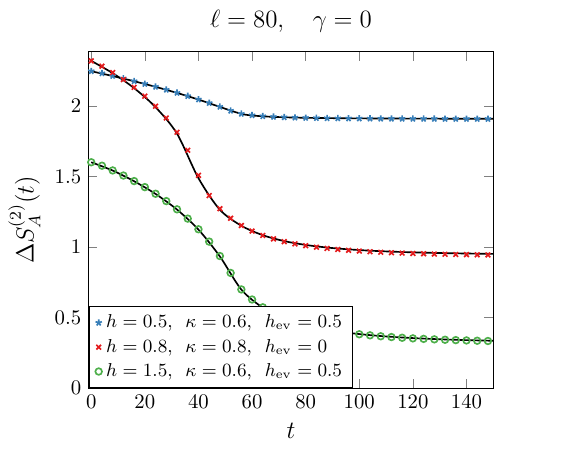}
    \caption{Dynamics of the second charge moment (left panel) and the second Rényi asymmetry (right panel) after a unitary quench ($\gamma=0$) from the ground state of the anisotropic SSH model. The data are shown for different parameters of the initial ($h$) and the evolution ($h_{\rm ev}$) magnetic fields and anisotropy $\kappa$. The solid curves in the left panel are obtained from \eqref{eq:quasiparticle_chargedmoments}, while those in the right panel by integrating \eqref{eq:quasiptclZ_n} over $\alpha$. The saturation of the curves in the right panel to non-zero values signals the lack of symmetry restoration under this dynamics.}
    \label{fig:2ndquench_gamma0}
\end{figure}

The expressions \eqref{eq:logZnsmalltimes} and \eqref{eq:logZn} can be directly inserted into the quasiparticle formulas to determine the evolution of the charged moments, and consequently, the Rényi asymmetries and Rényi entropies.
To derive the quasiparticle formula for the Rényi entropies, we evaluate $\mathcal{M}^{(n)}_{\boldsymbol{\alpha}}(k,t\to\infty)$ at $\boldsymbol{\alpha}=\boldsymbol{0}$, yielding
\begin{equation}
   S_A^{(n)}(t\to\infty)\sim\frac{\ell}{1-n}\int_{-\pi}^{\pi} \frac{dk}{8\pi}\min(2v_k t/\ell,1)\log \det \mathcal{M}^{(n)}_{\boldsymbol{0}}(k,t\to\infty)\,,
\end{equation}
where $v_k=\vert\varepsilon'_{k/2}\vert$ is the velocity of the quasiparticles with momentum $k$ for the unitary evolution considered in this protocol and $\varepsilon_k=\sqrt{(4+h_{\rm ev}^2+(4-h_{\rm ev}^2)\cos k)/2} $ is the dispersion relation.
The fact that the velocities are defined from the dispersion relation with halved momentum is due to the invariance under two-site translations of the initial state \eqref{eq:initialstate_k_2ndquench}, as highlighted in \cite{Ares:2023kcz}.

As for the evolution of the entanglement asymmetry, we first need to provide a quasiparticle formula for the ratio  $Z_n(\boldsymbol{\alpha},t)/Z_n(\boldsymbol{\alpha},t\to 0)$, which reads 
\begin{equation}
\label{eq:quasiparticle_chargedmoments}
\log\left(\frac{Z_n(\boldsymbol{\alpha},t)}{Z_n(\boldsymbol{\alpha},t\to 0)}\right)\sim
\frac{\ell}{4}\int_{-\pi}^{\pi} \frac{dk}{2\pi}\min(2v_k t/\ell,1)\log\left(\frac{ \det \mathcal{M}^{(n)}_{\boldsymbol{\alpha}}(k,t\to\infty)}{\det \mathcal{M}^{(n)}_{\boldsymbol{\alpha}}(k,t\to 0)}\right).
\end{equation}
In the left panel of Fig.\,\ref{fig:2ndquench_gamma0}, we benchmark this prediction for $n=2$ against numerical computations, finding a good agreement. 
When $\boldsymbol{\alpha}=0$, Eq. \eqref{eq:quasiparticle_chargedmoments} reduces to $(1-n)S_A^{(n)}(t)$. Thus, we can conveniently rewrite Eq. \eqref{eq:quasiparticle_chargedmoments} as
\begin{equation}
\label{eq:quasiptclZ_n}
Z_n(\boldsymbol{\alpha},t)=Z_n(\boldsymbol{0},t)
e^{\ell \left(A_n(\boldsymbol{\alpha})+ B_n(\boldsymbol{\alpha},t/\ell)+B'_n(\boldsymbol{\alpha},t/\ell)\right)} ,
\end{equation}
where 
\begin{equation}
  A_n(\boldsymbol{\alpha})=\int_{-\pi}^{\pi}\frac{d k}{8\pi}\log \det \mathcal{M}^{(n)}_{\boldsymbol{\alpha}}(k,t\to0)  ,
\end{equation}
\begin{equation}
   B_n(\boldsymbol{\alpha},t/\ell)=-\int_{-\pi}^{\pi}\frac{d k}{8\pi}\min(2v_k t/\ell,1)\log \det \mathcal{M}^{(n)}_{\boldsymbol{\alpha}}(k,t\to0) ,
\end{equation}
\begin{equation}
   B'_n(\boldsymbol{\alpha},t/\ell)= 
   \int_{-\pi}^{\pi}\frac{dk}{8\pi}\min(2v_k t/\ell,1)\log\left(\frac{\det \mathcal{M}^{(n)}_{\boldsymbol{\alpha}}(k,t\to\infty)}{\det \mathcal{M}^{(n)}_{\boldsymbol{0}}(k,t\to \infty)}\right).
\end{equation}
The fact that, at large times, $B'_n(\boldsymbol{\alpha},t/\ell)\neq 0$ and therefore $Z_n(\boldsymbol{\alpha},t)$ retains a dependence on $\boldsymbol{\alpha}$ implies that the dynamics studied in this section does not restore the initially broken $U(1)$ symmetry.
The formula \eqref{eq:quasiptclZ_n} can be plugged into \eqref{eq:FT} and used to compute the entanglement asymmetry, at least numerically, for any integer $n$.
The lack of symmetry restoration under this Hermitian dynamics is evident in the right panel of Fig.~\ref{fig:2ndquench_gamma0}, where the second Rényi asymmetries saturate to non-zero values at late times. We observe a good agreement between the numerical data and the analytic (black solid) curves obtained by integrating \eqref{eq:quasiptclZ_n} with $n=2$ over $\alpha$. %, according to \eqref{eq:FT}.

To better understand the dynamics of entanglement asymmetry analytically, we once again focus on the case $n=2$.
In the early time regime, we take the exponential of \eqref{eq:logZ2smalltimes}, we plug it in the integral \eqref{eq:FT} with $n=2$ and apply a saddle-point approximation, following the computation in \cite{Ares:2023kcz}. We obtain
\begin{equation}
\label{eq:asymmetry2smalltime}
\Delta S_A^{(2)}(t=0)=\frac{1}{2}\log \ell+\frac{1}{2}\log\frac{\pi g_0^{(2)}}{2}+O(\ell^{-1}),    
\end{equation}
where
\begin{equation}
\label{eq:g_0_2}
  g_0^{(2)}=\int_{-\pi}^{\pi}\frac{d k}{2\pi}16\vert u_{1,k}(0)\vert^2,
\end{equation}
and $u_{1,k}(0)$ is given in \eqref{eq:u1_timezero}.
As for the late time behavior, by adapting the same computation to the charged moments in Eq. \eqref{eq:quasiptclZ_2}, we obtain
\begin{equation}
\label{eq:asymmetry2latetime}
\Delta S_A^{(2)}(t\to\infty)=\frac{1}{2}\log \ell+\frac{1}{2}\log\frac{\pi g_\infty^{(2)}}{2}+O(\ell^{-1}),    
\end{equation}
where
\begin{equation}
\label{eq:g_infty_2}
  g_\infty^{(2)}=\int_{-\pi}^{\pi}\frac{d k}{2\pi}\frac{4\vert\zeta_\infty(k)\vert^2}{m^{(\infty)}_{0}(k)}.
\end{equation}
It is worth noticing than both $g_0^{(2)}$ and $g_\infty^{(2)}$ vanish when $\kappa\to 0$, i.e. when the symmetry is restored. As a consequence, both \eqref{eq:asymmetry2smalltime} and \eqref{eq:asymmetry2latetime} are ill-defined when $\kappa\to 0$. As discussed in \cite{Ares:2023kcz} for the entanglement asymmetry after a quench from the tilted Néel state, this is due to the non-commutativity between the limit $\ell\to \infty$ and the limit where the symmetry is not initially broken.

We remark that the physics that we observe in this (unitary) second quench, is very similar to the one found for the tilted Neel state: even though the evolution conserves the $U(1)$ symmetry, we believe that the breaking of the translation to a 2-site unit cell has the same effect here, i.e. it prevents the symmetry restoration. 

\textbf{The large $\kappa$ regime.---} 
 Before analyzing the behavior in the non-unitary setup, we make a remark about a possible scenario in which the symmetry is restored. From Eqs. \eqref{eq:asymmetry2latetime} and \eqref{eq:g_infty_2}, we observe that the late time asymmetry becomes ill-defined when $\kappa\to \infty$ and $h_{\rm ev}\to 0$. Since the same happens for $\kappa\to 0$, i.e. when the symmetry is explicitly restored in the ground state of Eq. \eqref{eq:Ham2}, this might imply that also in this regime the symmetry is dynamically restored by the evolution. To address this point more precisely, we take the $\kappa\to \infty$ limit of $g_\infty^{(2)}$, where its expression simplifies and the integrand in Eq. \eqref{eq:g_infty_2} can be written in terms of the original parameters of the quench, namely
 \begin{equation}
\label{eq:ginf_largekappa}
  g_\infty^{(2)}=\int_{-\pi}^{\pi}\frac{d k}{2\pi}\frac{8 h_{\rm ev}^2\vert \sin(k/2)\vert^2\left[4+h_{\rm ev}^2-(h_{\rm ev}^2-4)\cos k\right]}{\left[2 h_{\rm ev}^2\vert \sin(k/2)\vert^2+4+h_{\rm ev}^2-(h_{\rm ev}^2-4)\cos k\right]^2}.
\end{equation}
This expression shows immediately that $g_\infty^{(2)}=0$ when $h_{\rm ev}= 0$, leading to the aforementioned divergence of the late time asymmetry. Moreover, we remark the limit $\kappa\to \infty$ has washed out the dependence of $g_\infty^{(2)}$ and $\Delta S_A^{(2)}(t\to\infty)$ on the parameter $h$ of the initial state.
Moving to the early time regime, we observe that $g_0^{(2)}\to 4$ as $\kappa\to \infty$ and therefore the asymmetry at time $t=0$ does not retain any dependence on the initial state in this regime.

\subsubsection{Non-Hermitian dynamics and Mpemba effect}

\begin{figure}[t!]
\hspace{-0.6cm}
    \includegraphics[width=0.54\textwidth]{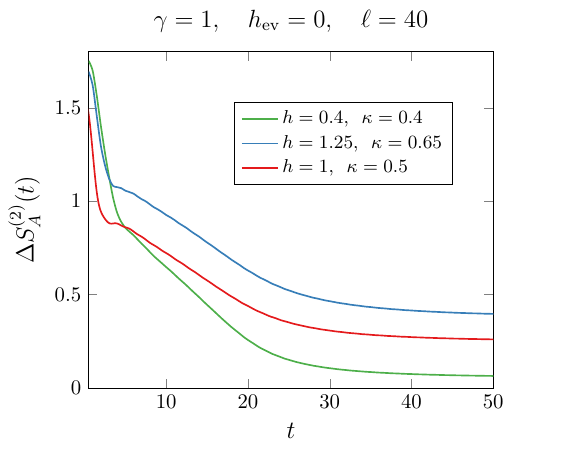}
\includegraphics[width=0.54\textwidth]{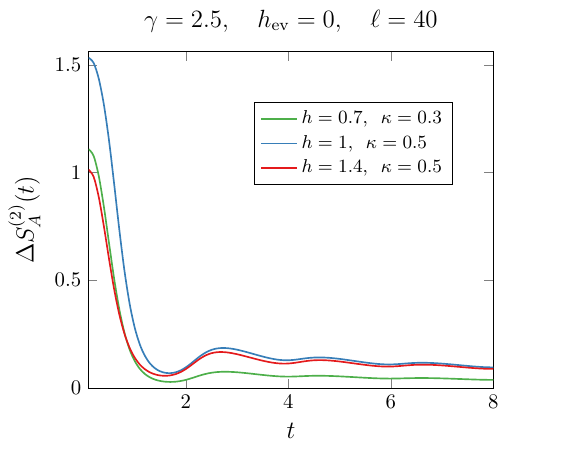}
    \caption{Dynamics of the entanglement asymmetry $\Delta S_A^{(2)}$ after the quench from the anisotropic SSH ground state. The data are shown for $h_{\rm ev}=0$ and  different parameters $\gamma< 4$, magnetic field $h$ and anisotropy $\kappa$. 
    As witnessed by the entanglement asymmetry not decaying to zero, the symmetry is not dynamically restored for the protocols considered in this figure.
    }
    \label{fig:2ndquench_Asymm2_nobreak}
\end{figure}

Now, we turn to the evolution under the Hamiltonian \eqref{eq:Ham2_ev} with $\gamma\neq 0$. 
Given the similarity between the quench setup studied here and that in Ref.~\cite{Caceffo:2024jbc}, we anticipate that the non-unitary dynamics will resemble a quench from an antiferromagnetic state with dissipative gain/loss terms. 
Indeed, Ref.~\cite{Caceffo:2024jbc} demonstrated that even an infinitesimal amount of dissipation in that setup leads to symmetry restoration. We now investigate whether the same holds in our case. 
Compared to the unitary case, symmetry restoration occurs if the terms $ \zeta(k,t)$  in the symbol \eqref{eq:Symbol_2nd_quench} vanish. This happens when the dispersion relation $\varepsilon_k=\sqrt{(4+h_{\rm ev}^2+(4-h_{\rm ev}^2)\cos k)/2-\gamma^2}$ becomes purely imaginary, i.e. when $(4+h_{\rm ev}^2+(4-h_{\rm ev}^2)\cos k)/2-\gamma^2\leq 0$. 
This inequality is satisfied for $\gamma\geq 4$, for any value of $h_{\rm ev}$. As a consequence, the symmetry can be restored only for large values of the measurement rate $\gamma$. 
Furthermore, we observe that the time average of the terms $\upsilon(k,t)$ in Eq. \eqref{eq:Symbol_2nd_quench}, which correspond to the correlators $\braket{c^{\dagger}_{k_o}c^{\dagger}_{k_o}}$, behaves differently depending on $\gamma$. For weak measurements $\gamma<4$, these terms remain nonzero, whereas at strong measurement rate $\gamma\geq 4$, they decay exponentially to 0. 
This further confirms that the symmetry is restored for $\gamma\geq 4$, since all the correlators responsible for the $U(1)$ symmetry breaking in Eq. \eqref{eq:Symbol_2nd_quench} vanish. 
These results outline a genuine \textit{measurement-induced symmetry restoration}, a key result of our work. 
In Fig.\,\ref{fig:2ndquench_Asymm2_nobreak}, we show that the second Rényi asymmetry does not decay to zero when $\gamma< 4$. This means that the $U(1)$ symmetry is not dynamically restored in this regime of parameters. Notice that the larger $\gamma$, the smaller the late-time value of the entanglement asymmetry, which ultimately becomes zero when $\gamma=4$. 

\begin{figure}[t!]
\hspace{-0.6cm}
    \includegraphics[width=0.54\textwidth]{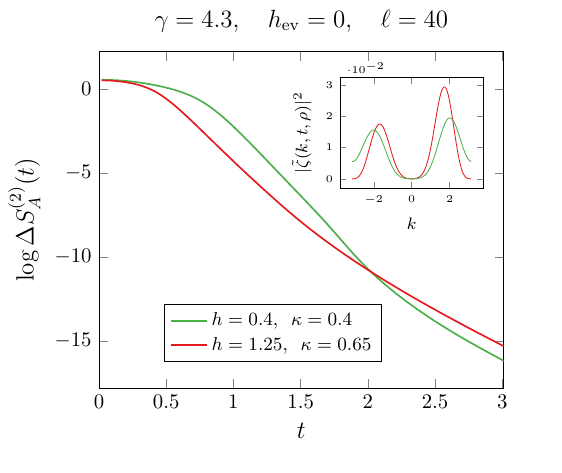}
     \includegraphics[width=0.54\textwidth]{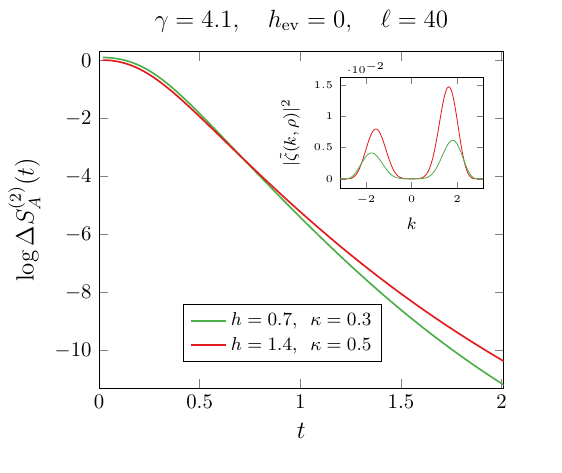}
    \caption{Dynamics of the entanglement asymmetry $\Delta S_A^{(2)}$ after the quench from the anisotropic SSH ground state. The data are shown for $h_{\rm ev}=0$ and  different parameters $\gamma>4$, magnetic field $h$ and anisotropy $\kappa$. As shown in the insets, the occurrence of the quantum Mpemba effect is established by the relations between the functions $\tilde{\zeta}$ in \eqref{eq:crit} evaluated for the two states.}
    \label{fig:2ndquench_Asymm2}
\end{figure}

Having identified a critical value of $ \gamma $ at which symmetry is restored, the next key question is whether certain choices of the initial parameters $ h $ and  $\kappa$  give rise to the quantum Mpemba effect. 
Although we lack an analytical prediction for the late-time behavior of the entanglement asymmetry, we can adopt a criterion similar to the one discussed for a quench starting from the antiferromagnetic state and evolving under a $U(1) $-symmetric Hamiltonian, as explored in Ref.~\cite{Caceffo:2024jbc}.
In that setup, the crossing in the entanglement asymmetry occurs if initial states with larger asymmetry lead
to pairs $\braket{c^{\dagger}_kc^{\dagger}_{\pi-k}}$ with lower \textit{asymmetry content}, that is quantified by $\int_{-\pi}^{\pi}\frac{dk}{2\pi}|\braket{c^{\dagger}_kc^{\dagger}_{\pi-k}}|^2$.
In our case, it means that the quantum Mpemba effect can arise if the more asymmetric initial state has a smaller value of $\int_{-\pi}^{\pi}\frac{dk}{2\pi}|\braket{c^{\dagger}_{k_o}c^{\dagger}_{k_e}}|^2$, since $\braket{c^{\dagger}_{k_o}c^{\dagger}_{k_e}}$ is responsible for the symmetry breaking at $\gamma<4$. We check this condition in Fig.\,\ref{fig:2ndquench_Asymm2}, where we observe that, given two initial states, $\rho_1$ and $\rho_2$, with $ \Delta S_A^{(n)}(\rho_1,t=0)>\Delta S_A^{(n)}(\rho_2,t=0)$, then for large time we find that 
\begin{equation}\label{eq:crit}
    \Delta S_A^{(n)}(\rho_1,t)<\Delta S_A^{(n)}(\rho_2,t) \Leftrightarrow  \int_{-\pi}^{\pi}\frac{dk}{2\pi}|\tilde{\zeta}(k,\rho_1)|^2 > \int_{-\pi}^{\pi}\frac{dk}{2\pi}|\tilde{\zeta}(k,\rho_2)|^2,
\end{equation}
where $\tilde{\zeta}$ has been obtained from Eq. \eqref{eq:def_zeta} in the limit $e^{t\sqrt{\gamma^2-(4+h_{\rm ev}^2+(4-h_{\rm ev}^2)\cos k)/2}}\gg 1$. 
%In Fig.\,\ref{fig:2ndquench_Asymm2}, we verify the criterion \eqref{eq:crit} in some examples for the second Rényi asymmetry. 
%We observe that the occurrence of the quantum Mpemba effect is associated with the validity of the condition on the right-hand side of \eqref{eq:crit}, as shown in the insets.

\section{Discussion}
\label{sec:futurepersp}

We have investigated the evolution of entanglement entropies and entanglement asymmetry following global quenches driven by non-Hermitian Hamiltonians. Since these dynamics arise in the no-click limit of continuously monitored systems, our analysis provides a stepping stone for a comprehensive understanding of Mpemba phenomenology in measurement-induced evolution. 
Concretely, we considered two global quench protocols in one-dimensional spin systems, utilizing a fermionic representation. In both cases, the initial state explicitly breaks $U(1)$ symmetry, while the evolution Hamiltonians preserve it. This setup allowed us to examine whether and how the symmetry is dynamically restored under non-unitary evolution. Given the absence of a quasiparticle description in non-Hermitian dynamics, we analyzed the late-time behavior of correlation functions, entanglement entropies, and entanglement asymmetry to characterize the system’s relaxation.

In the first quench protocol, where the system evolves from the $XY$ ground state under a non-Hermitian $XX$ Hamiltonian, we found that entanglement entropies initially grow but then decay exponentially to zero, with a rate proportional to the non-Hermitian term. This differs from previous results with a different initial state, emphasizing the strong dependence of non-unitary dynamics on initial conditions. The $U(1)$ symmetry is ultimately restored, and crucially, the quantum Mpemba effect—when previously absent in the unitary case—can emerge even for small non-Hermitian contributions. We proposed that this effect can be explained through the density of Cooper pairs responsible for symmetry breaking, a conjecture supported by our numerical analysis.

In the second quench protocol, starting from the ground state of a modified SSH model, we observed a fundamentally different behavior. In the unitary case, the entanglement asymmetry saturates to a nonzero value, indicating a lack of symmetry restoration. However, introducing a non-Hermitian dimerized term leads to a critical threshold beyond which the $U(1)$ symmetry is dynamically restored, signaling a genuine measurement-induced phenomenon. In this regime, the quantum Mpemba effect also appears, and we again verified its connection to Cooper pair densities.

Our results highlight the critical role of non-unitary dynamics in entanglement evolution and symmetry restoration, opening several key research avenues. 
First, overcoming the no-click limit requires a more refined analysis. For quadratic systems the recently developed replica field theories could offer a deeper, universal characterization of Mpemba effects in monitored fermionic systems~\cite{buchhold2021effective,fava2024monitored,fava2023nonlinear,jian2023measurementinducedentanglementtransitionsquantum,jian2022criticalityentanglementnonunitaryquantum,muller2025monitoredinteractingdiracfermions,poboiko2023theory,chahine2023entanglement,poboiko2025measurement,leung2024entanglementoperatorcorrelationsignatures,klocke2024powerlaw,klocke2023majorana}. 
On a broader perspective, throughout this work, we have interpreted the presence, or absence, of the Mpemba effect in terms of the density of correlation functions responsible for symmetry breaking. While this interpretation aligns with our findings, we aim to establish a more rigorous criterion that can predict, a priori, when the Mpemba effect can occur in monitored systems. 
These ideas immediately connect with the effect of quench disorder in the Hamiltonian of monitored systems, where the interplay between monitoring and localization can drastically alter the entanglement growth~\cite{PhysRevB.108.165126,PhysRevResearch.2.043072,piccitto2025entanglementbehaviorlocalizationproperties,tomasi2024stable}. 
Whether disorder suppresses or enhances the quantum Mpemba effect remains an open question, especially in light of recent studies on disordered many-body systems \cite{liu2024quantummpembaeffectsmanybody,Joshi:2024sup}. 
Toward interacting systems, quantum circuits with $U(1)$ symmetry~\cite{agrawal2022entanglementandcharge} as well as interacting non-Hermitian Hamiltonians provide an important direction. Complementarily, the study of monitored long-range systems can lead to analytically amenable results for the measurement-induced symmetry restoration and the Mpemba effect ~\cite{russomanno2023entanglement,passarelli2024many,delmonte2024measurementinducedphasetransitionsmonitored,10.21468/SciPostPhysCore.5.2.023,minato2022fate,PhysRevLett.128.010604,muller2022measurementinduced,piccitto2023entanglement}.
Furthermore, linking the Mpemba effect in open quantum systems to the measurement-induced phenomena studied here is an important open question. Partially monitored many-body systems \cite{ladewig2022monitoredopenfermion,PhysRevB.106.024304,PhysRevB.109.144306,salatino2024exploringtopologicalboundaryeffects} provide a promising framework for quantitative predictions. We leave these explorations for future work.

\authorcontributions{
GDG and SM performed the analytical calculations. 
XT provided the code for the numerical simulations.
GDG and SM equally contributed to the writing. 
All authors codesigned the project. 
}

\funding{
G.D.G is supported by the ERC Consolidator grant (number: 101125449/acronym: QComplexity).  Views and opinions expressed are however those of the authors only and do not necessarily reflect those of the European Union or the European Research Council. Neither the European Union nor the granting authority can be held responsible for them.
X.T. acknowledges DFG under Germany's Excellence Strategy – Cluster of Excellence Matter and Light for Quantum Computing (ML4Q) EXC 2004/1 – 390534769, and DFG Collaborative Research Center (CRC) 183 Project No. 277101999 - project B01. }

% \dataavailability{The code for the numerical simulations and the data are available at~\cite{code}.} 

\acknowledgments{We thank Filiberto Ares and Pasquale Calabrese for their comments about the draft. SM thanks the support from the Walter Burke Institute for Theoretical Physics and the Institute for Quantum Information and Matter at Caltech, where most of this work has been elaborated.
}

\conflictsofinterest{The authors declare no conflicts of interest.} 

%%%%%%%%%%%%%%%%%%%%%%%%%%%%%%%%%%%%%%%%%%
%% Optional

%\newpage
%%%%%%%%%%%%%%%%%%%%%%%%%%%%%%%%%%%%%%%%%%
%% Optional
\appendixtitles{no} 
\appendixstart
\appendix

\section[\appendixname~\thesection]{Time evolution of the SSH ground state with anisotropic terms}
\label{app:Details}

In this appendix, we provide computational details on the quench dynamics described in Sec.\,\ref{sec:2ndprotocol}.
The first step is to find the explicit expressions of the coefficients $u_{j,k}(0)$, which determine the fixed-momentum contribution \eqref{eq:initialstate_k_2ndquench} to the initial ground state.
Adapting the procedure in Sec.\,\ref{subsec:correlation_1stquench}, we introduce a six-dimensional representation of the action of $H_k(h,\kappa)$ in \eqref{eq:Ham2} on $\vert\psi_k(0)\rangle$
\begin{equation}
\label{eq:6drep_Hamiltonian}
 H_k(h,\kappa) \vert \psi_k(0)\rangle\quad \Leftrightarrow \quad  
 \mathcal{M}_k(h,\kappa,0) \begin{pmatrix}
         u_{1,k}(0)
        \\
      u_{2,k}(0)
       \\
      u_{3,k}(0)
       \\
      u_{4,k}(0)
       \\
      u_{5,k}(0)
       \\
      u_{6,k}(0)
    \end{pmatrix}\,,
\end{equation}
where
\begin{equation}
\label{eq:Ham_6d}
\mathcal{M}_k(h,\kappa,\gamma)=\begin{pmatrix}
       0  & 0  & 0  & 0  &B_{k}(\kappa) &-B_{-k}(\kappa)
        \\
         0  & -i\gamma  & 0  & 0  &A_{k}(h) &A_{-k}(h)
         \\
     0  & 0  & i\gamma  & 0  &A_{k}(h) &A_{-k}(h)
      \\
     0  & 0  & 0  & 0  &B_{k}(\kappa) &-B_{-k}(\kappa)
      \\
      B_{-k}(\kappa) & A_{-k}(h)  & A_{-k}(h)  & B_{-k}(\kappa)  &0 &0
      \\
      -B_{k}(\kappa) & A_{k}(h)  &A_{k}(h)   & -B_{k}(\kappa)  &0 &0
    \end{pmatrix},
\end{equation}
and $A_k(h)$ and $B_k(\kappa)$ are given by \eqref{eq:AandB_def}.
This is because, as discussed in the main text, the fixed-$k$ contributions to the ground state are expanded in terms of six non-vanishing independent combinations of fermionic bilinears.
The expressions of $u_{j,k}(0)$ as functions of the parameters $k$, $h$ and $\kappa$ are given by the components of the normalized eigenvector of $\mathcal{M}_k(h,\kappa,0)$ associated with the smallest eigenvalue. 
We can work them out and we obtain
\begin{equation}\label{eq:u1_timezero}
 u_{1,k}(0) = u_{4,k}(0)=\frac{i}{\sqrt{2}}\frac{{\rm Im}[A_k(h)B_{-k}(\kappa)]\sqrt{\vert A_k(h)\vert^2+\vert B_k(h,\kappa)\vert^2+\vert F_k(h,\kappa)\vert}}{A_{-k}(h)F_k(h,\kappa)+A_{k}(h)\vert F_k(h,\kappa)\vert} \,, 
\end{equation}
\begin{eqnarray}
   &&u_{2,k}(0) = u_{3,k}(0)=
   \\
   \nonumber
  && \frac{-\sqrt{\left(\vert A_k(h)\vert^2+\vert B_k(h,\kappa)\vert^2+\vert F_k(h,\kappa)\vert\right)\left[A_k(h)^2B_{-k}(\kappa)+B_{k}(\kappa)(\left\vert F_k(h,\kappa)\vert-\vert B_k(h,\kappa)\vert^2\right)\right]}}{4\sqrt{2}{\rm Re}[A_k(h)B_{-k}(\kappa)]F_k(h,\kappa)}\,, 
\end{eqnarray}
\begin{equation}
   u_{5,k}(0) =\frac{1}{2}\frac{\vert F_k(h,\kappa)\vert}{F_k(h,\kappa)} \,, \quad
  u_{6,k}(0) =\frac{1}{2} \,,
\end{equation}
where, to make the notation more compact, we have defined the combination
\begin{equation}
F_k(h,\kappa)\equiv  A_k(h)^2 -B_k(\kappa)^2\,.
\end{equation}
The explicit expressions of $u_{j,k}(0)$ in terms of the parameters $h,\kappa$ are cumbersome and not insightful. 
%Thus, we do not report them here.

Once the initial ground state is identified, we let it evolve through the Hamiltonian \eqref{eq:Ham2_ev}. The evolved state at time $t>0$ has the form \eqref{eq:evolvedstate_2ndquench}. To determine the time dependence of the six coefficients, we solve the Schroedinger equation with the evolution Hamiltonian written in the six-dimensional representation introduced above. This corresponds to the matrix \eqref{eq:Ham_6d} with $h=h_{\rm ev}$, $\kappa=0$  and $\gamma\neq 0$. The initial conditions of the Schroedinger equation are given by the parameters $u_{j,k}(0)$ determined before.
The solutions of the resulting system of differential equations can be worked out analytically, but their expression is cumbersome and we do not report here. In the case of a unitary quench, i. e. $\gamma=0$, the expressions simplify a bit and we find
\begin{eqnarray}
 \label{eq:u1t_2ndquench}
 u_{1,k}(t)&=&
 u_{1,k}(0)
 \,,
 \\
u_{2,k}(t) &=&
u_{3,k}(t) =
\frac{1}{2\varepsilon_k}
\bigg[2\varepsilon_k\big(\cos(2t\varepsilon_k)u_{2,k}(0)\big)-
\\
\nonumber
&&
i\sin(2t\varepsilon_k)\big(A_k(h_{\rm ev})u_{5,k}(0)+A^*_k(h_{\rm ev})u_{6,k}(0)\big)
\bigg]
 \,,
 \\
u_{4,k}(t) &=&
u_{4,k}(0)
 \,,
 \\
u_{5,k}(t) &=&
\frac{A^*_k(h_{\rm ev})}{\varepsilon_k^3}\bigg[A_k(h_{\rm ev})\varepsilon_k u_{5,k}(0)\cos^2\left(t\varepsilon_k\right)-
\\
\nonumber
&&
A^*_k(h_{\rm ev})\left(\varepsilon_k u_{6,k}(0)\sin^2\left(t\varepsilon_k\right)+i\sin(2t\varepsilon_k)A_k(h_{\rm ev})u_{2,k}(0)\right)
\bigg]
 \,,
 \\
 \label{eq:u6t_2ndquench}
u_{6,k}(t) &=&
\frac{A_k(h_{\rm ev})}{\varepsilon_k^3}\bigg[-A_k(h_{\rm ev})\varepsilon_k u_{5,k}(0)\sin^2\left(t\varepsilon_k\right)+
\\
\nonumber
&&
A^*_k(h_{\rm ev})\left(\varepsilon_k u_{6,k}(0)\cos^2\left(t\varepsilon_k\right)-i\sin(2t\varepsilon_k)A_k(h_{\rm ev})u_{2,k}(0)\right)
\bigg]
 \,,
\end{eqnarray}
where 
\begin{equation}
\label{eq:dispersion_2ndquench}
   \varepsilon_k\equiv\vert A_k(h_{\rm ev})\vert=\sqrt{(4+h_{\rm ev}^2+(4-h_{\rm ev}^2)\cos k)/2}\,.
\end{equation}
Once plugged into \eqref{eq:evolvedstate_2ndquench}, \eqref{eq:u1t_2ndquench}-\eqref{eq:u6t_2ndquench} give the evolution of the initial state under the unitary case of the quench protocol considered in Sec.\,\ref{sec:2ndprotocol}. Moreover, these coefficients can be used in \eqref{eq:def_xi}-\eqref{eq:def_zeta} to determine the evolution of the two-point functions along the considered dynamics.

\section[\appendixname~\thesection]{Useful properties of Toeplitz matrices}
\label{app:Toeplitz}

In this appendix, we report some formulas for the determinants of Toeplitz matrices, which can be exploited to derive \eqref{eq:logZnsmalltimes} and \eqref{eq:logZn} in the main text.

A Toeplitz matrix can be in general written as
\begin{equation}
\label{eq:Toeplitz_def}
  (T_\ell[g(k)])_{jj'}=\int_{-\pi}^{\pi}\frac{d k}{2\pi}e^{-i k(j-j')}g(k) ,\qquad j,j'=1,\dots,\ell,
\end{equation}
where the $d\times d$ matrix $g(k)$ is called symbol and is defined for $k\in[0,2\pi)$. The size of the matrix in \eqref{eq:Toeplitz_def} is therefore $d\cdot \ell$.  Notice that the correlation matrices \eqref{eq:Gammawithsymbol} and \eqref{eq:corr_time t} are Toeplitz matrices with $2\times 2$ and $4\times 4$ symbols respectively. In \cite{Ares:2023kcz}, the following conjectures on the asymptotic behaviours of determinants involving Toeplitz matrices have been formulated. 
Given the product of $N$ Toeplitz matrices, the first conjecture says that, for large $\ell$,
 \begin{eqnarray}
 \label{eq:conjecture1}
    \log\det\left[I+\prod_{m=1}^N T_\ell[g_m(k)]\right] \sim A\ell,
 \end{eqnarray}
 where
 \begin{eqnarray}
 \label{eq:conjecture1_coeff}
     A\equiv\int_{-\pi}^{\pi}\frac{d k}{2\pi}\log\det\left[I+\prod_{m=1}^N g_m(k)\right],
 \end{eqnarray}
provided that $\det\left[I+\prod_{m=1}^N g_m(k)\right]\neq 0$.
Setting $n=2$ in \eqref{eq:numerics} and recalling that the correlation matrices and $e^{i\alpha n_A}$ are Toeplitz matrices, we find that $\log Z_2(\alpha)$ can be written as the left-hand side of \eqref{eq:conjecture1}. Thus, we can use \eqref{eq:conjecture1} and \eqref{eq:conjecture1_coeff} to compute the $n=2$ charge moments in the limit $\ell\to\infty$.

The second conjecture we report here concerns products of Toeplitz matrices and their inverses. Given a set of Toeplitz matrices $T_\ell[g_m(k)]$ and a set of inverse Toeplitz matrices $T_\ell[g'_m(k)]^{-1}$ with $m=1,\dots,N$, if we assume that the symbols $g'_m(k)$ are invertible, then we have
 \begin{eqnarray}
 \label{eq:conjecture2}
    \log\det\left[I+\prod_{m=1}^N T_\ell[g_m(k)]T_\ell[g'_m(k)]^{-1}\right] \sim A'\ell,
 \end{eqnarray}
 where
 \begin{eqnarray}
 \label{eq:conjecture2_coeff}
     A'\equiv\int_{-\pi}^{\pi}\frac{d k}{2\pi}\log\det\left[I+\prod_{m=1}^N g_m(k)g'_m(k)^{-1}\right].
 \end{eqnarray}
We can again notice that, since the correlation matrices and $e^{i\alpha n_A}$ are Toeplitz matrices, the formula \eqref{eq:numerics} for the charge moments can be written using \eqref{eq:conjecture2} and, therefore,  \eqref{eq:conjecture2_coeff} allows to determine the asymptotic behaviour at large subsystem size.
In particular, using \eqref{eq:conjecture2} and  \eqref{eq:conjecture2_coeff}, we have derived the expressions \eqref{eq:logZnsmalltimes} and \eqref{eq:logZn} in the main text.

\begin{adjustwidth}{-\extralength}{0cm}

\reftitle{References}


\bibliography{newbib}

% \PublishersNote{}
\end{adjustwidth}
\end{document}